# Estimating the Size of *Aedes aegypti* Populations from Dengue Incidence Data: Implications for the risk of Yellow Fever Outbreaks

Eduardo Massad[1,2,3,*], Marcos Amaku[1], Francisco Antonio Bezerra Coutinho[1], Claudio José Struchiner[4], Luis Fernandez Lopez[1], Annelies Wilder-Smith[5,6,7,8] and Marcelo Nascimento Burattini[1,9]

[1]School of Medicine, University of Sao Paulo, Brazil.

[2]London School of Hygiene and Tropical Medicine, UK.

[3]College of Life and Natural Sciences, The University of Derby, UK.

[4]Programme of Scientific Computation, Fundação Oswaldo Cruz, Rio de Janeiro, Brazil.

[5]Li Ka Shing Knowledge Institute, St Michael's Hospital, Toronto, Canada.

[6]Institute of Public Health, University of Heidelberg, Germany.

[7]Department Public Health and Clinical Medicine, Epidemiology and Global Health, Umeå University, SE-901 85 Umeå, Sweden.

[8]Lee Kong Chian School of Medicine, Nanyang Technological University, Singapore.

[9]Hospital São Paulo, Escola Paulista de Medicina, Universidade Federal de São Paulo, São Paulo, SP, Brazil.
*Correspondence to: edmassad@usp.br

**Abstract**

In this paper we present a model to estimate the density of aedes mosquitoes in a community affected by dengue. The method consists in fitting a continuous function to the incidence of dengue infections, from which the density of infected mosquitoes is derived straightforwardly. Further derivations allow the calculation of the latent and susceptible mosquitoes' densities, the sum of the three equals the total mosquitoes' density. The method is illustrated with the case of the risk of urban yellow fever resurgence in dengue infested areas but the same procedures apply for other aedes-transmitted infections like Zika and chikungunya viruses.

## 1. Introduction

The main components of the Ross-Macdonald model for vector-borne infections have been estimated with reasonable degree of accuracy (Coutinho et al., 2006; Massad et al., 2011; Massad & Coutinho 2012; Amaku et al., 2013, 2016; Lopez et al., 2016).

Values for mosquitoes' biting and mortality rates, extrinsic incubation periods, probabilities of transmission from mosquitoes-to-humans and vice-versa, human recovery and mortality rates from infection, are found in the specialized literature (Liu-Helmerson et al., 2016). Mosquitoes' densities, however, vary from place to place and with time and are extremely difficult to estimate (Adams and Kaplan, 2009). Empirical efforts (Maciel-de-Freitas, Eiras, and Lourenço de Oliveira, 2008) to determine the actual size of the mosquitoes populations in affected areas are limited in number and very laboriously done. Even these are limited in space and time due to regional and seasonal variations.

*Aedes aegypti* is known to transmit several infections like dengue virus, yellow fever virus, chikungunya virus, and Zika virus (ECDC, 2017). Some authors (Larsen & Ashley, 1971) suggested to be a potential vector of Venezuelan Equine Encephalitis virus and vector competency studies have shown *Ae. aegypti* is capable of transmitting West Nile virus. West Nile virus has also been isolated from this mosquito species in the field (Turell et al., 2005).

In this paper we propose a method to indirectly estimate the density of aedes mosquitoes in dengue affected areas. The method is based on incidence data of dengue infections and an application is illustrated with the case of the risk of urban yellow fever in a dengue infested area of Brazil. It serves, however, for the estimation of the risk of any aedes-transmitted disease outbreak, like Zika virus, chikungunya, Mayaro, among others (ECDC, 2017).

## 2. Materials and Methods

*2.1.Formalism*
*2.1.1.The Ross-Macdonald Model*

We use a variant of the classical Ross-Macdonald model, described in details in (Coutinho et al., 2006; Amaku et al., 2015, 2016).

The populations involved in the transmission are human hosts and mosquitoes. Therefore, the population densities are divided into the following compartments: susceptible humans denoted $S_H$; infected humans, $I_H$; recovered (and immune) humans, $R_H$; total humans, $N_H$; susceptible mosquitoes, $S_M$; infected and latent mosquitoes, $L_M$; infected and infectious mosquitoes, $I_M$. The variables and parameters appearing in the model are defined in Tables 1 and 2, respectively.

The model is defined by the following equations:

$$\frac{dS_H}{dt} = -abI_M \frac{S_H}{N_H} + \mu_H (N_H - S_H)$$

$$\frac{dI_H}{dt} = abI_M \frac{S_H}{N_H} - (\mu_H + \gamma_H) I_H$$

$$\frac{dR_H}{dt} = \gamma_H I_H - \mu_H R_H$$

$$\frac{dS_M}{dt} = -acS_M \frac{I_H}{N_H} + \mu_M (L_M + I_M) + \frac{dN_M}{dt}$$

$$\frac{dL_M}{dt} = acS_M \frac{I_H}{N_H} - \gamma_M L_M - \mu_M L_M$$

$$\frac{dI_M}{dt} = \gamma_M L_M - \mu_M I_M$$

$$\frac{dN_M}{dt} = \begin{cases} 0 \text{ for constant population size} \\ \text{or} \\ \Omega \cos(2\pi f t + \varphi) N_M(t) \text{ for seasonal variation} \end{cases}$$

$$N_H = S_H + I_H + R_H$$
$$N_M = S_M + L_M + I_M$$
(1)

*Remark 1: This model differs from the classical Ross-Macdonald model because the extrinsic incubation period in the classical Ross-Macdonald model is assumed to last $\tau$ days, whereas in model (1) we assumed an exponential distribution for the latency in the*

*mosquitoes. The classical Ross-Macdonald model can be obtained from system (1) by replacing the fifth and sixth equations by* (Amaku et al., 2014):

$$\frac{dL_M}{dt} = acS_M \frac{I_H}{N_H} - \mu_M L_M - acS_M(t-\tau)\frac{I_H(t-\tau)}{N_H(t-\tau)}e^{-\mu_M \tau}$$

$$\frac{dI_M}{dt} = acS_M(t-\tau)\frac{I_H(t-\tau)}{N_H(t-\tau)}e^{-\mu_M \tau} - \mu_M I_M$$

*where $\tau$ is the extrinsic incubation period and $\mu_M$ is the mosquito mortality rate. The expressions developed below in this paper with equations (1) can be replaced by the corresponding expressions of the classical Ross-Macdonald model described above by replacing $\frac{\gamma_M}{\gamma_M + \mu_M}$ by $e^{-\mu_M \tau}$. $\gamma_M$ is related to $\tau$ by $\tau = -\frac{1}{\mu_M}\ln\left[\frac{\gamma_M}{\gamma_M + \mu_M}\right]$.*

*The form of the extrinsic incubation period is not known experimentally, to the best of our knowledge. Both assumptions mentioned above are therefore arbitrary. We choose the exponential decay of latency because it simplifies the calculations.*

First we identify a dengue outbreak. For the purposes of this paper, an outbreak is defined as beginning at the moment $t_i$ when the epidemic curve is at its lowest values, that is, when $\frac{d}{dt}I_H(t_i) = 0$. The outbreak ends at time $t_f$, when $\frac{d}{dt}I_H(t_f) = 0$ again.

*2.1.2. Calculating Dengue incidence from notification data in a population previously unexposed to dengue viruses*

Second, having identified a dengue outbreak, we fitted a continuous function to the number of actually reported dengue cases multiplied by 4 to take into account the 4:1 asymptomatic:symptomatic ratio (Ximenes et al., 2016), which has the form:

$$Incidence_{DENV}(t) = c_1 \exp\left[-\frac{(t-c_2)^2}{c_3}\right] + c_4 \qquad (2)$$

representing the time-dependent dengue infection incidence. In equation (2), $c_1$ is a scale parameter that determines the maximum incidence, $c_2$ is the time at which the maximum incidence is reached, $c_3$ represents the width of the time-dependent incidence function and $c_4$ is just another scaling parameters. Equation (2) is intended to reproduce a "Gaussian" curve and so $c_1$ and $c_4$ are just scale parameters but $c_2$ represents the "mean" (and mode or maximum) time and $c_3$ represents the "variance" of the time distribution of cases. All parameters $c_i, i = 1,...,4$ were fitted to model (2) so that the force of infection when applied to the dynamical model described below reproduces the observed incidence of dengue for a given outbreak in a region preferably small.

The first term of the first equation in system of equations (1) models the number of new infections per time unit. In terms of the classical notation of vector-borne infections (Coutinho et al., 2006), it is equal to the product of the force-of-infection, $\lambda(t)$ times the number of susceptible humans, denoted $S_H(t)$. As is well known, the force-of-infection in vector-borne infections is the product of the biting rate times the probability of transmission from infected mosquitoes to the human hosts, times the number of infected mosquitoes divided by the total number of humans (Coutinho et al., 2006). In terms of the variables of the model, the force of infection is defined as follows.

Let $S_H$ and $I_H$ represent the susceptible and infected humans, respectively, and $\lambda(t)$ be the force of infection (or incidence-density rate) which, as mentioned above, represents the product of the mosquitoes biting rate, $a$, the probability of transmission from mosquitoes to humans, $b$, and the number of infected mosquitoes with respect to humans, $\frac{I_M}{N_H}$, and is normally denoted $\lambda = ab\frac{I_M}{N_H}$. As mentioned in the subheading of this section, all individuals in this population are considered susceptible to dengue, that is, $S_H(0) = N_H$.

*Remark 2: Note that* $\lambda(t)S_H(t)$ *is the dengue incidence. Or in detail,*

$$Incidence(t) = \lambda(t)S_H(t) = ab\frac{I_M(t)}{N_H}S_H(t).$$

By numerically adjusting model (2) to the actual data we found the values of the parameters $c_i, i = 1,...,4$ that generate $\lambda(t)S_H(t)$, that is, the incidence data (reported cases); in other words, the fitted function $Incidence_{DENV}(t)$ (equation (2)) is used to the system of equations (1) in order to check the incidence $\lambda(t)S_H(t)$.

The fitted incidence (taking into account the 4:1 asymptomatic:symptomatic ratio (Ximenes et al., 2016)), for two neighborhoods of the city of Rio de Janeiro in 2011-2012, using the parameters values shown in Table 2, are shown in Figs. 1 and 2.

## 3. Calculating the density of mosquitoes from the incidence of a dengue outbreak

In order to calculate the density of mosquitoes, we shall need the expression and derivatives of the incidence estimated (fitted) in the previous section.

*3.1 Calculating the derivatives of* $Incidence_{DENV}(t)$

From equation (2), we have:

$$\frac{d}{dt}Incidence_{DENV}(t) = -2c_1\left[\frac{(t-c_2)}{c_3}\right]\exp\left\{-\frac{(t-c_2)^2}{c_3}\right\} \quad (3)$$

$$\frac{d^2}{dt^2}Incidence_{DENV}(t) = -\frac{2c_1}{c_3}\left\{1-\left[\frac{2(t-c_2)^2}{c_3}\right]\right\}\exp\left\{-\frac{(t-c_2)^2}{c_3}\right\} \quad (4)$$

$$\frac{d^3}{dt^3}Incidence_{DENV}(t) = \frac{4c_1}{c_3^2}(t-c_2)\left[3-\frac{2(t-c_2)^2}{c_3}\right]\exp\left[-\frac{(t-c_2)^2}{c_3}\right] \quad (5)$$

*3.2 Calculating the derivatives of* $S_H(t)$

From the first equation of the Ross-Macdonald model for the susceptible humans we obtain:

$$\frac{d}{dt}S_H(t) = -ab\frac{I_M(t)}{N_H}S(t) + \mu_H(N_H - S_H(t)) \qquad (6)$$

But, as mentioned above, $Incidence_{DENV}(t) = \lambda(t)S_H(t) = ab\frac{I_M(t)}{N_H}S(t)$. Hence:

$$\frac{d}{dt}S_H(t) = -Incidence_{DENV}(t) + \mu_H(N_H - S_H(t)) \qquad (7)$$

Therefore:

$$S_H(t) = N_H[1-\exp(-\mu_H t)] + \exp(-\mu_H t)\left\{S_H(0) - \int_0^t \exp(-\mu_H s)Incidence_{DENV}(s)ds\right\} \qquad (8)$$

or, in terms of the parameters of equation (2):

$$S_H(t) = e^{-\mu_H t}S_H(0) - c_4\frac{(1-e^{-\mu_H t})}{\mu_H} + N_H(1-e^{-\mu_H t}) - \frac{c_1\sqrt{c_3\pi}}{2}e^{\left[c_3\left(\frac{c_2}{c_3}-\frac{\mu_H}{2}\right)^2 - \frac{c_2^2}{c_3}\right]}e^{-\mu_H t} \times$$
$$\times \left\{erf\left[\sqrt{\frac{1}{c_3}}t + \sqrt{c_3}\left(-\frac{c_2}{c_3}+\frac{\mu_H}{2}\right)\right] - erf\left[\sqrt{c_3}\left(-\frac{c_2}{c_3}+\frac{\mu_H}{2}\right)\right]\right\} \qquad (9)$$

where $erf(x) = \frac{1}{\sqrt{\pi}}\int_0^x e^{-t^2}dt$ is the error function.

For the numerical simulations, we used $S_H(0) = N_H$. This is consistent with the case when the population do not have a history of previous exposure to dengue.

Hence:

$$\frac{d^2}{dt^2}S_H(t) = -\frac{d}{dt}Incidence_{DENV}(t) - \mu_H\frac{d}{dt}S_H(t) \qquad (10)$$

$$\frac{d^3}{dt^3}S_H(t) = -\frac{d^2}{dt^2}Incidence_{DENV}(t) - \mu_H\frac{d^2}{dt^2}S_H(t) \qquad (11)$$

*3.3 Calculating the derivatives of $I_H(t)$*

From the second equation of the Ross-Macdonald model for the infected humans we obtain:

$$\frac{d}{dt}I_H(t) = ab\frac{I_M(t)}{N_H}S(t) - (\mu_H + \gamma_H)I_H(t) = Incidence_{DENV}(t) - (\mu_H + \gamma_H)I_H(t) \quad (12)$$

which can be solved by standard methods resulting in:

$$I_H(t) = I_H(0)e^{-(\mu_H+\gamma_H)t} + \int_0^t e^{(\mu_H+\gamma_H)(x-t)} Incidence_{DENV}(x)dx \quad (13)$$

For the numerical simulations, we used $I_H(0) = \dfrac{Incidence_{DENV}(0)}{(\mu_H + \gamma_H)}$, which follows from equation (12) and implies $\dfrac{d}{dt}I_H(0) = 0$. Note that when the population do not have a history of previous exposure to dengue, $Incidence_{DENV}(0) = 1$, meaning that one case was introduced in the population.

*3.4 Calculating the number of mosquitoes $N_M(t) = S_M(t) + L_M(t) + I_M(t)$*

*3.4.1 Infective mosquitoes $I_M(t)$*

We know that:

$$Incidence_{DENV}(t) = \lambda(t)S_H(t) = ab\frac{I_M(t)}{N_H}S(t) \quad (14)$$

Therefore:

$$I_M(t) = \frac{N_H}{S_H(t)}\frac{Incidence_{DENV}(t)}{ab} \quad (15)$$

$$\frac{d}{dt}I_M(t) = \frac{N_H}{S_H(t)}\frac{1}{ab}\left[\frac{d}{dt}Incidence_{DENV}(t) - \frac{Incidence_{DENV}(t)}{S_H(t)}\frac{d}{dt}S_H(t)\right] \quad (16)$$

$$\frac{d^2}{dt^2} I_M(t) = \frac{N_H}{S_H(t)} \frac{1}{ab} \left\{ \frac{2 \text{Incidence}_{DENV}(t)}{(S_H(t))^2} \left( \frac{d}{dt} S_H(t) \right)^2 - \frac{\text{Incidence}_{DENV}(t)}{S_H(t)} \frac{d^2}{dt^2} S_H(t) \right.$$
$$\left. - \frac{2}{S_H(t)} \frac{d}{dt} S_H(t) \frac{d}{dt} \text{Incidence}_{DENV}(t) + \frac{d^2}{dt^2} \text{Incidence}_{DENV}(t) \right\}$$

(17)

$$\frac{d^3}{dt^3} I_M(t) = \frac{N_H}{S_H(t)} \frac{1}{ab} \left\{ -\frac{6}{(S_H(t))^3} \left( \frac{d}{dt} S_H(t) \right)^3 \text{Incidence}_{DENV}(t) \right.$$
$$+ \frac{5}{(S_H(t))^2} \left( \frac{d}{dt} S_H(t) \right)^2 \frac{d}{dt} \text{Incidence}_{DENV}(t)$$
$$+ \frac{4}{(S_H(t))^2} \left( \frac{d}{dt} S_H(t) \right) \left( \frac{d^2}{dt^2} S_H(t) \right) \text{Incidence}_{DENV}(t)$$
$$- \frac{3}{(S_H(t))} \frac{d^2}{dt^2} S_H(t) \frac{d}{dt} \text{Incidence}_{DENV}(t)$$
$$- \frac{3}{(S_H(t))} \frac{d}{dt} S_H(t) \frac{d^2}{dt^2} \text{Incidence}_{DENV}(t)$$
$$\left. - \frac{\text{Incidence}_{DENV}(t)}{(S_H(t))} \frac{d^3}{dt^3} S_H(t) + \frac{d^3}{dt^3} \text{Incidence}_{DENV}(t) \right\}$$

(18)

*3.4.2 Latent mosquitoes $L_M(t)$*

From the sixth equation of the Ross-Macdonald model for the infected mosquitoes we obtain:

$$\frac{d}{dt} I_M(t) = \gamma_M L_M(t) - \mu_M I_M(t) \tag{19}$$

Therefore:

$$L_M(t) = \frac{1}{\gamma_M} \left[ \frac{d}{dt} I_M(t) + \mu_M I_M(t) \right] \tag{20}$$

Hence:

$$\frac{d}{dt} L_M(t) = \frac{1}{\gamma_M} \left[ \frac{d^2}{dt^2} I_M(t) + \mu_M \frac{d}{dt} I_M(t) \right] \tag{21}$$

$$\frac{d^2}{dt^2} L_M(t) = \frac{1}{\gamma_M} \left[ \frac{d^3}{dt^3} I_M(t) + \mu_M \frac{d^2}{dt^2} I_M(t) \right] \qquad (22)$$

*3.4.3 Susceptible mosquitoes $S_M(t)$*

From the fifth equation of the Ross-Macdonald model for latent mosquitoes we obtain:

$$\frac{d}{dt} L_M(t) = ac \frac{I_H(t)}{N_H} S_M(t) - (\mu_M + \gamma_M) L_M(t) \qquad (23)$$

Therefore:

$$S_M(t) = \frac{N_H}{ac I_H(t)} \left[ \frac{d}{dt} L_M(t) + (\mu_M + \gamma_M) L_M(t) \right] \qquad (24)$$

Hence:

$$\frac{d}{dt} S_M(t) = \left\{ \frac{N_H}{ac I_H(t)} \left[ \frac{d^2}{dt^2} L_M(t) + (\mu_M + \gamma_M) \frac{d}{dt} L_M(t) \right] \right. \\
\left. - \frac{N_H}{ac I_H^2(t)} \frac{d}{dt} I_H(t) \left[ \frac{d}{dt} L_M(t) + (\mu_M + \gamma_M) L_M(t) \right] \right\} \qquad (25)$$

The total size of the mosquitoes population, $N_M(t)$ is given by:

$$N_M(t) = S_M(t) + L_M(t) + I_M(t) \qquad (26)$$

or

$$N_M(t) = \left\{ \frac{N_H}{ac I_H(t)} \left[ \frac{d}{dt} L_M(t) + (\mu_M + \gamma_M) L_M(t) \right] + \frac{1}{\gamma_M} \left[ \frac{d}{dt} I_M(t) + \mu_M I_M(t) \right] \right. \\
\left. + \frac{N_H}{S_H(t)} \frac{Incidence_{DENV}(t)}{ab} \right\} \qquad (27)$$

## 4. Illustrating the method

*4.1. Example of applications*

*4.1.1.Testing the model's experimental consistency*

In order to test the model's accuracy, we applied the formalism above to the borough of Olaria in Rio de Janeiro. Olaria is located at the north of the city of Rio de Janeiro and is a traditional suburban residential area of the city. In the 2000 census, Olaria had an estimated population of 62,509 inhabitants in an area of around 3.7 km$^2$. This borough was chosen because in 2007 Maciel-de-Freitas, Eiras and Lourenço-de-Oliveira (2008) carried out a study in the area, in which they estimated, through the MosquiTrap and aspirator method, the population of *Aedes aegyptii*. In the estimated 0.79 km$^2$ area covering the average flight range of aedes mosquitoes, the authors found 3,505 and 4,828 female mosquitoes in the MosquiTrap and aspirator, respectively, totalizing 8,333.

Using the data from dengue in the same period, the model estimated a total aedes population in the 0.79 km$^2$ area of Olaria in a period of two weeks as 8,145±12 female mosquitoes, which is a good approximation to the empirical data.

*4.1.2. The case of dengue in two other neighborhoods of Rio de Janeiro*

After fitting the dengue incidence in a given outbreak for a specific region, we used the above formalism to calculate the total mosquito density by simulating system (1). For this, we need, in addition to the parameters values as in Table 2, the initial condition for the susceptible mosquitoes, $S_M(0)$.

- *Botafogo*

Botafogo is a beachfront neighborhood of the city of Rio de Janeiro, Southeastern Brazil. It is essentially an upper middle class with small commerce community, with a population of about 83000 people, distributed in an area of 479.90 ha.
We used dengue data for the period between October 2011 and December 2012, comprising 3140 infections. Figure 1 shows the fitting of equation (2) to the monthly number of dengue infections in Botafogo. The parameters values used in the calculations are shown in Table 2.

- *São Cristóvão*

São Cristóvão is a traditional neighborhood located in the Central area of Rio de Janeiro, Brazil. With a population of about 26000 people distributed in an area of 410.56 ha, São Cristóvão experienced 3248 dengue infections in the period between October 2011 and December 2012.

Figure 2 shows the fitting of equation (2) to the monthly number of dengue infections in São Cristóvão.

In Fig. 3 we show the result of the calculation of the total number of Aedes mosquitoes in both neighborhoods, using the parameters as in Table 2.

- *Combining the data from both neighborhoods*

In this section, we show that the method can be used for small geographical areas where the infection transits by mosquitoes' movements but can also be applied for larger aggregated areas, where the infection transits by infected humans movements.

Let us call the incidence in areas 1 and 2 as $Incidence_1(t)$ and $Incidence_2(t)$, respectively, and defined as:

$$Incidence_1(t) = abI_{M_1}(t)\frac{S_{H_1}(t)}{N_{H_1}(t)},$$

and  (28)

$$Incidence_2(t) = abI_{M_2}(t)\frac{S_{H_2}(t)}{N_{H_2}(t)}$$

$$Incidence_{1+2}(t) = ab\left[I_{M_1}(t)\frac{S_{H_1}(t)}{N_{H_1}(t)} + I_{M_2}(t)\frac{S_{H_2}(t)}{N_{H_2}(t)}\right] \quad (29)$$

If we define:

$$I_{M_1}(t) = qI_M(t),$$
$$I_{M_2}(t) = (1-q)I_M(t),$$
$$S_{H_1}(t) = pS_H(t),$$
$$S_{H_2}(t) = (1-p)S_H(t), \qquad (30)$$
$$N_{H_1}(t) = pN_H(t), \text{ and}$$
$$N_{H_2}(t) = (1-p)N_H(t),$$

then:

$$Incidence_{1+2}(t) = ab\left[qI_M(t)\frac{pS_H(t)}{pN_H(t)} + (1-q)I_M(t)\frac{(1-p)S_H(t)}{(1-p)N_H(t)}\right] \qquad (31)$$

and the fractions $q, (1-q), p$ and $(1-p)$ cancel out, reducing equation (31) to equation (14).

*Remark 3: About the above calculation, one should note that: (1) the values of p and q can be time-dependent; (2) the formalism can be extended to any number of sites. Note, however, that by combining sites we lose the spatial distribution of mosquitoes. We get only the total number.*

Figure 4 illustrates this reasoning for the neighborhoods of Botafogo and São Cristóvão combined.

Note that the numerical simulation is a good approximation of the sum of the number of mosquitoes of each borough.

## 5. Calculating Dengue incidence from notification data in a population previously exposed to dengue viruses

The only modification necessary for this case from the previous discussed formalism occurs when we test the model's consistency. When the population has been previously exposed to dengue, the boundary conditions must be modified.
In this case only a fraction $p$ of the human population is susceptible to dengue due to previous epidemics, that is $S_H(0) \to pN_H - I_H(0)$ and $R_H(0) \to (1-p)N_H$ in the initial conditions of system (1). The implications of this is as follows.

First, consider the Effective Reproduction Number, $R_{\text{eff}}(t)$ (Massad and Coutinho 2012) of system (1):

$$R_{\text{eff}}(t) = \frac{a^2 bc \gamma_m N_m(t)}{\mu_M(\mu_H + \gamma_H)(\mu_M + \gamma_M)N_H} \frac{p S_H^*(t)}{N_H} \tag{32}$$

where $S_H^*(0) = N_H$. There is a threshold $p_{th}$ that makes $R_{\text{eff}}(t) < 1$ for $t > 0$. When $p \to p_{th}$, then it is necessary a larger mosquitoes population to explain the same number of infections observed. When $p < p_{th}$, then it is impossible to have an outbreak in these places and the formalism breaks down. When $p = 1$, then we have a minimum mosquitoes population that reproduces the observed number of cases. In contrast, when $p \to p_{th}$, the mosquitoes population tends to its maximum size. This maximum size is calculated using the expression of $R_0 = 1$ ($R_0$ is $R_{\text{eff}}(t=0)$). Therefore, in the case where the population has been previously exposed to dengue, the total size of the susceptible mosquitoes population (density) is given by $p S_H^*(t)$ ($0 \le p \le 1$).

## 6. Digging a little bit more on the methodology proposed: testing the model's theoretical consistency

In this section, we examine how the method proposed above deals with an artificially constructed outbreak. To artificially construct an outbreak of a hypothetical vector-borne infection we specify a function $N_M(t)$ (see below) and use it in a conventional Ross-Macdonald model.

We know (Coutinho et al., 2006; Amaku et al., 2015, 2016) that a pure Ross-Macdonald model usually produces a single outbreak with a narrow and high peak in the incidence of cases (later we show an exception). In nature, these narrow and high peak are seldom observed because, as explained in (Amaku et al., 2016) the outbreak is produced in waves, that is, the disease travels throughout a geographical area. The Ross-Macdonald model only reproduces an outbreak of this type if we concentrate on data from an area small enough (of the order of the area covering few times the mosquitoes' flight range).

We produced three pure Ross-Macdonald models, one with a constant mosquito population and two in which the mosquito population oscillates with time (the incidence in one of these last outbreaks is bi-modal with time). In the three cases, not surprisingly, we discovered that we could not fit the outbreak with the 'Gaussian' type of function as in equation (2). The resulting fit was always broad and the mosquito's population retrieved was in poor agreement with the artificial input. We tried to solve this problem by replacing equation (2) with (31):

$$Incidence = c_1 \text{sech}^2(c_2 t + c_3) + c_4 \qquad (33)$$

but the fitted outbreak was not good in the case of the one-modal outbreak and, of course, very poor for the bi-modal outbreak

With these artificially created outbreaks, we used a different approach to retrieve the mosquito population that, in this case, we pretend not to know. First we calculated, from the artificially constructed incidence, the values of $S_H(t)$ from equation (8) and $I_M(t)$ from equation (15). Then, by numerically differentiating, when necessary, from equations (16-27), we calculated the values of $L_M(t)$ and $S_M(t)$. Next we checked the above calculations by using the value of the artificially constructed incidence to calculate the human prevalence $I_H(t)$ as in equation (13) and then $I_M(t)$, $L_M(t)$ and $S_M(t)$ by solving the differential equations of the Ross-Macdonald model (1). Note that, as mentioned above, this approach is not suitable to be applied to natural outbreaks, unless the data from outbreak is obtained from a very limited geographical area, where the infection transits by infected mosquitoes movements.

In Figs. 5 (*a* and *b*) we show the incidence of one artificially constructed outbreak assuming a constant mosquitoes population (see Fig. 5a) and the retrieved number of mosquitoes populations using only the generated outbreak's incidence and compare it with the number of mosquitoes generated by the Ross-Macdonald model (see Fig. 5b). As can be seen, the agreement is almost perfect.

In Figs. 6 (a and b) we show the incidence of an artificially constructed outbreak assuming an oscillating mosquitoes population according to equation (34).

$$\frac{dN_M(t)}{dt} = \Omega \cos(2\pi f t + \varphi) N_M(t) \tag{34}$$

In Fig. 6a we show the generated artificially outbreak and in Fig 6b we show the retrieved number of mosquitoes populations using only the generated outbreak's incidence and compare it with the number of mosquitoes generated by a Ross-Macdonald model, assuming an oscillating mosquitoes population according to equation (34). As can be seen, the agreement is almost perfect.

In Fig 7a we show a bi-modal outbreak generated by a oscillating population of mosquitoes as in equation (34). The bi-modal outbreak is obtained by using a different set of initial conditions. Finally in Fig 7b we show the retrieved number of mosquitoes populations using only the generated outbreak's incidence and compare it with the number of mosquitoes generated by a Ross-Macdonald model. Again in this case the agreement is almost perfect.

**7. Estimating the risk of urban yellow fever resurgence in dengue endemic cities**

In this section, we calculate the risk of yellow fever resurgence and the expected number of autochthonous cases in the neighborhoods of Botafogo and São Cristóvão analysed in section 4.1.2. This risk was calculated assuming an infected traveler, arriving at each month of the year in any one of those neighborhoods.

First we used the number of mosquitoes from dengue incidence from equation (27), described in section 4.1.2 for the dengue season of 2011-2012. Then we used the Ross-Macdonald model, described below, with the parameters related to yellow fever, denoted by the subscript *yf* and as initial conditions for the susceptible individuals the respective populations of these Rio's neighborhoods. The model has the form:

$$\frac{dS_H}{dt} = (-ab_{yf}I_M \frac{S_H}{N_H} + \mu_H(N_H - S_H))\theta(t - t_0)$$

$$\frac{dI_H}{dt} = (ab_{yf}I_M \frac{S_H}{N_H} - (\mu_H + \gamma_{H_{yf}} + \alpha_{H_{yf}})I_H)\theta(t - t_0)$$

$$\frac{dR_H}{dt} = \gamma_{H_{yf}}I_H - \mu_H R_H$$

$$\frac{dS_M}{dt} = -ac_{yf}S_M \frac{I_H}{N_H} + \mu_M(L_M + I_M) + \frac{dN_M}{dt}$$

$$\frac{dL_M}{dt} = ac_{yf}S_M \frac{I_H}{N_H} - \gamma_{M_{yf}}L_M - \mu_M L_M$$

$$\frac{dI_M}{dt} = \gamma_{M_{yf}}L_M - \mu_M I_M$$

$$N_H = S_H + I_H + R_H$$
$$N_M = S_M + L_M + I_M$$

(35)

where $\theta(t - t_0)$ is the Heaviside equation and simulates the arrival of the infected traveler at $t = t_0$. $\frac{dN_M}{dt}$ is the sum of equations (16), (21) and (25). As mentioned before $S_H(0)$ is assumed to be the whole population of each neighborhood and $I_H(0) = 1$.

*Remark 4: $S_H(t)$, $I_H(t)$ and $R_H(t)$ are densities (Amaku et al., 2015). Therefore, to assume $I_H(0) = 1$ is to assume that a number of infected travelers invade the neighborhood and that their densities is 1 individual per unit area. This is unimportant if the area is small enough.*

For the neighborhood of Botafogo, the maximum number of autochthonous cases is reached when the imported infection arrives at around 7 months after October 2011, with the number of yellow fever infections peaking between 5 and 11 and serious cases peaking between 1 and 2 (Fig. 7).

For the neighborhood of São Cristóvão, the maximum number of autochthonous cases is reached when the imported infection arrives at around 4 months after October 2011, with the number of yellow fever infections peaking between 5 and 11 and serious cases peaking between 1 and 2 (Fig. 8).

To complete the above analysis, we calculated the probability that one infected traveler arriving in February 2012would generate at least one autochthonous yellow fever case.

As mentioned above, the risk of urban yellow fever resurgence depends on the size of the *Aedes* mosquitoes population and its vectorial competence. As explained in the main text, this is defined as the relative reduction in the parameters $c$ and $b$ specific for yellow fever with respect to those specific for dengue. Hence, for instance, we used the value 0.6 for both parameters in the case of dengue and multiplied $c$ and $b$ for yellow fever by a factor varying from 0 to 1. Note that we assumed that the local *Aedes* mosquitoes are always more competent to transmit dengue than yellow fever (Massad et al., 2001).

We then calculated:
1) the risk of yellow fever introduction (the probability of at least one autochthonous cases in the first generation of infective travelers) by one infective traveler to the neighborhoods of Rio de Janeiro arriving in February 2012. We remind that there was a huge outbreak of dengue in this dengue year of 2011-2012; and
2) the expected number of YF infections in the worst scenario after one year, that is, when the traveler arrives in the month of February 2012, both as a function of the local *Aedes* vector competence.

## 8. Discussion and Conclusions

In this paper we present a model to estimate the density of *Aedes* mosquitoes in a community affected by dengue. The model is based on the fitting of a continuous function to the incidence of dengue infections, from which the density of infected mosquitoes is derived straightforwardly. Further derivations allows the calculation of the latent and susceptible mosquitoes' densities, the sum of the three equals the total mosquitoes' density. The model is illustrated with the case of the risk of urban yellow fever resurgence in dengue infested areas but the same methods apply for other *Aedes*-transmitted infections like Zika and chikungunya viruses.

The model demonstrated to be reliable as the example of the Olaria neighborhood shows. It retrieved the actual number of mosquitoes collected in the area with good accuracy.

One caveat is worth noting; the Ross-Macdonald model assumes homogenously mixing population. Therefore, introducing one infected individual means to introduce a density of infected individuals that is homogeneously distribute over the whole area (Amaku et al., 2016). Therefore, the smaller the area we apply the model, the more reliable the results are.

The conclusion of the above analysis is that there is a positive and non-negligible risk of urban yellow fever resurgence in some dengue endemic areas due to their high *Aedes* mosquitoes densities. The actual risk will be dependent on the probability that at least one infective human arrives at the right moment of the year, that is, when the local population of aedes mosquitoes is increasing in size and also on their vector competence. The examples provided in this paper are only intended to illustrate the method and more accurate parameters estimations are necessary for the true estimation of the risk of resurgence of urban yellow fever in those areas infested by *Aedes aegypti*. Finally, estimating the risk of urban yellow fever resurgence is central for the designing of an optimum vaccination strategy due to the yellow fever vaccine adverse events (Massad et al., 2005).

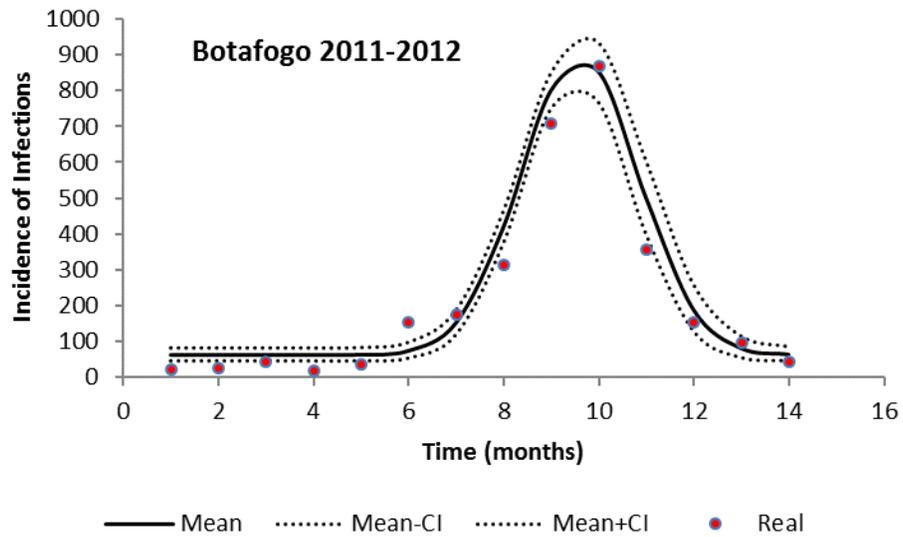

**Fig.1. Fitting a continuous function to the incidence of dengue infection in the period between October 2011 and December 2012 in Botafogo, Rio de Janeiro. Dots represent the actual notified data (x 4, see main text), continuous line the mean incidence and dotted line the 95% Confidence Interval.**

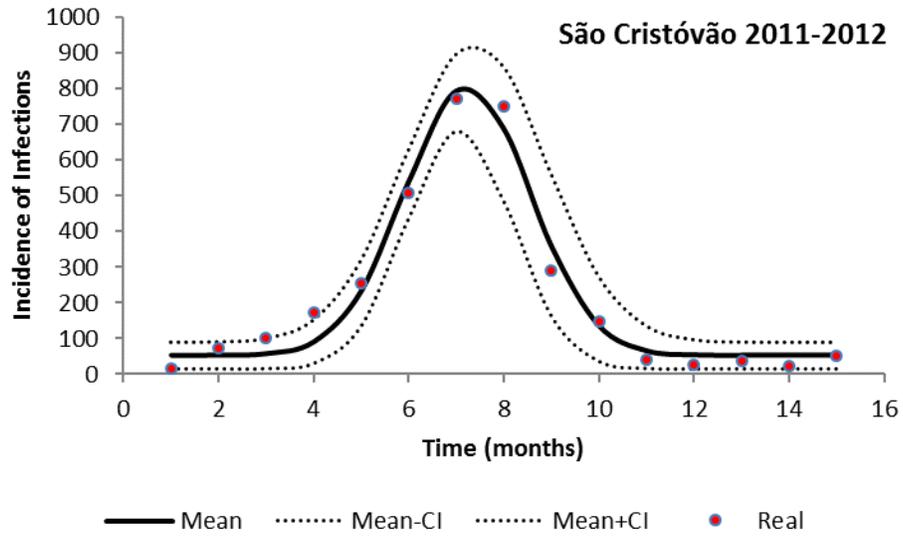

**Fig. 2.** Fitting a continuous function to the incidence of dengue infection in the period between October 2011 and December 2012 in São Cristóvão, Rio de Janeiro. Dots represent the actual notified data (x 4, see main text), continuous line the mean incidence and dotted line the 95% Confidence Interval.

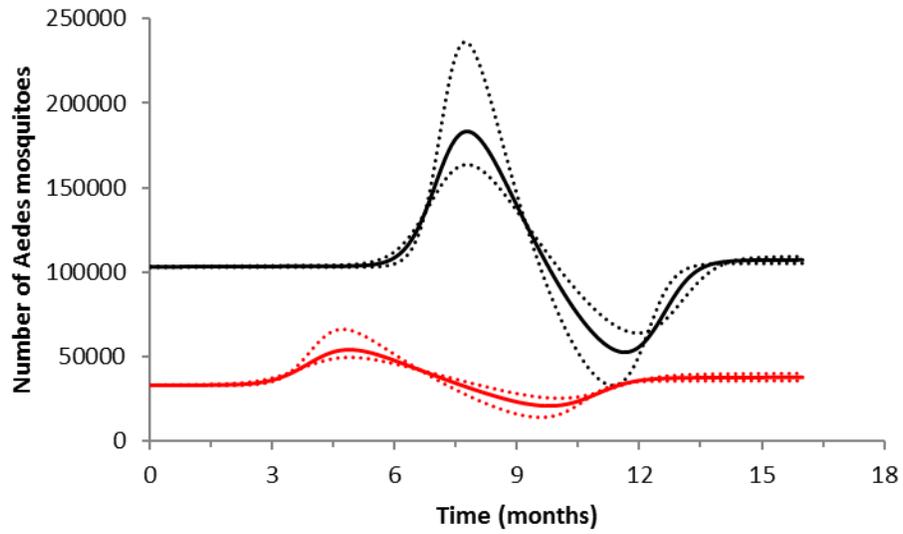

**Fig. 3.** Estimation of the size of the *Aedes* mosquitoes' population from the incidence of dengue infection in the period between October 2011 and December 2012 in Botafogo (red lines) and São Cristóvão (black lines) Rio de Janeiro. Continuous line the mean mosquitoes' population size and dotted line the 95% Confidence Interval.

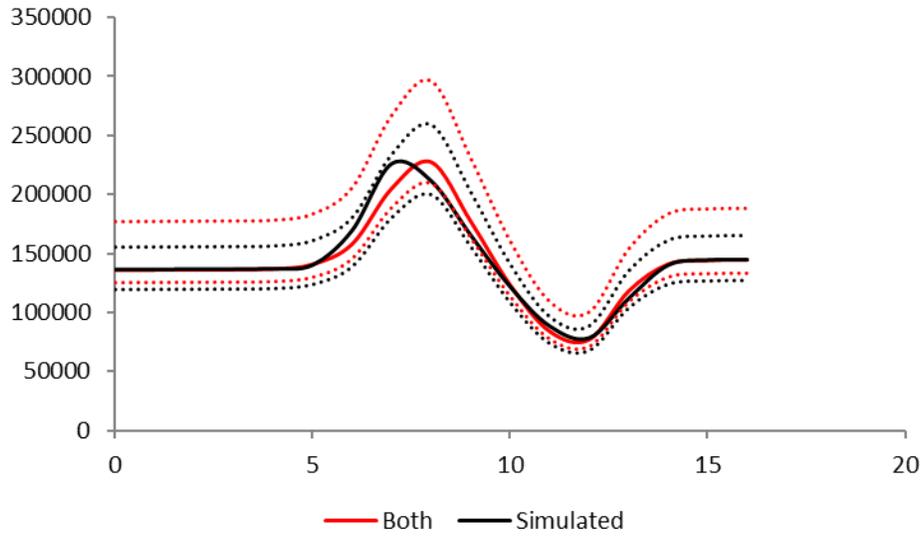

**Fig. 4.** Estimation of the size of the *Aedes* mosquitoes' population from the incidence of dengue infection in the period between October 2011 and December 2012 in Botafogo and São Cristóvão. Black lines represent the sum of both neighborhoods and red line the combination of them. Continuous line the mean mosquitoes' population size and dotted line the 95% Confidence Interval.

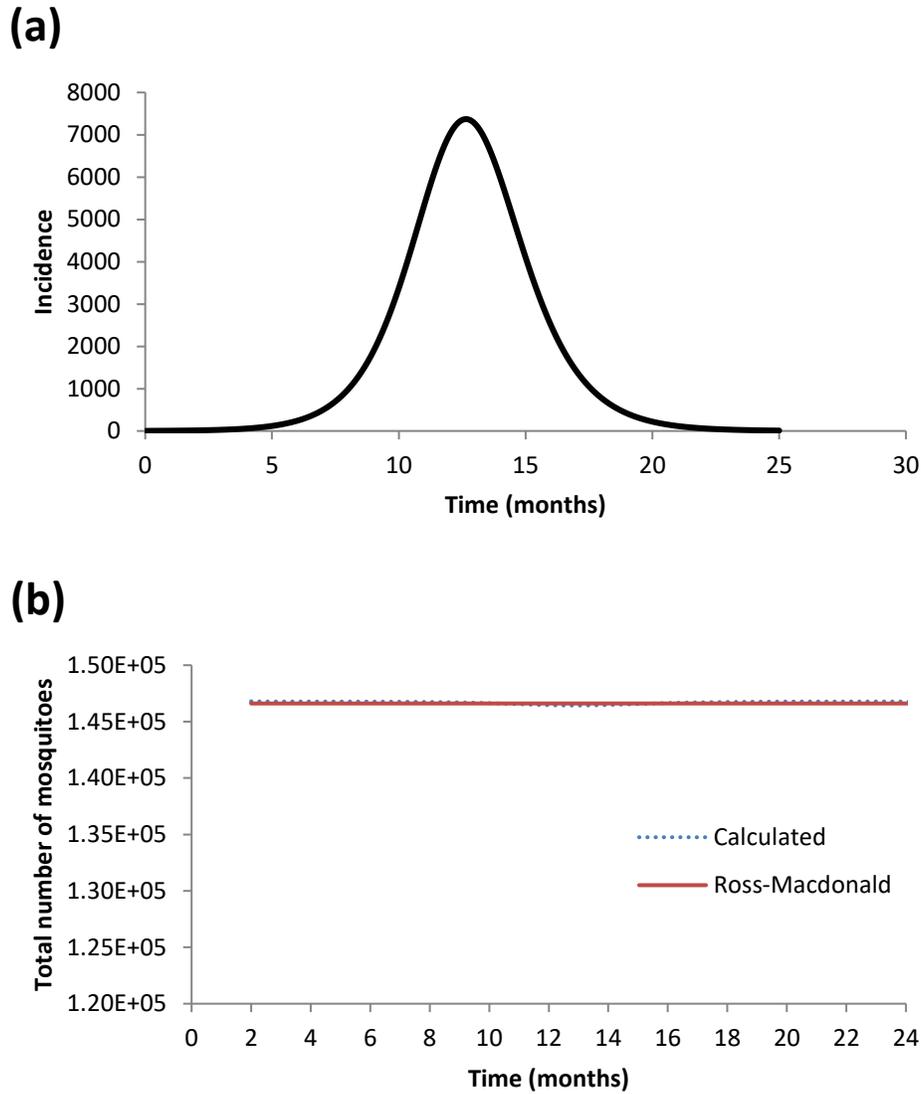

**Fig.5. (a) Dengue incidence outbreak constructed with a constant mosquito population. (b) Calculated number of mosquito populations (blue line) compared with that generated by a Ross-Macdonald model assuming a constant mosquito population (red line).**

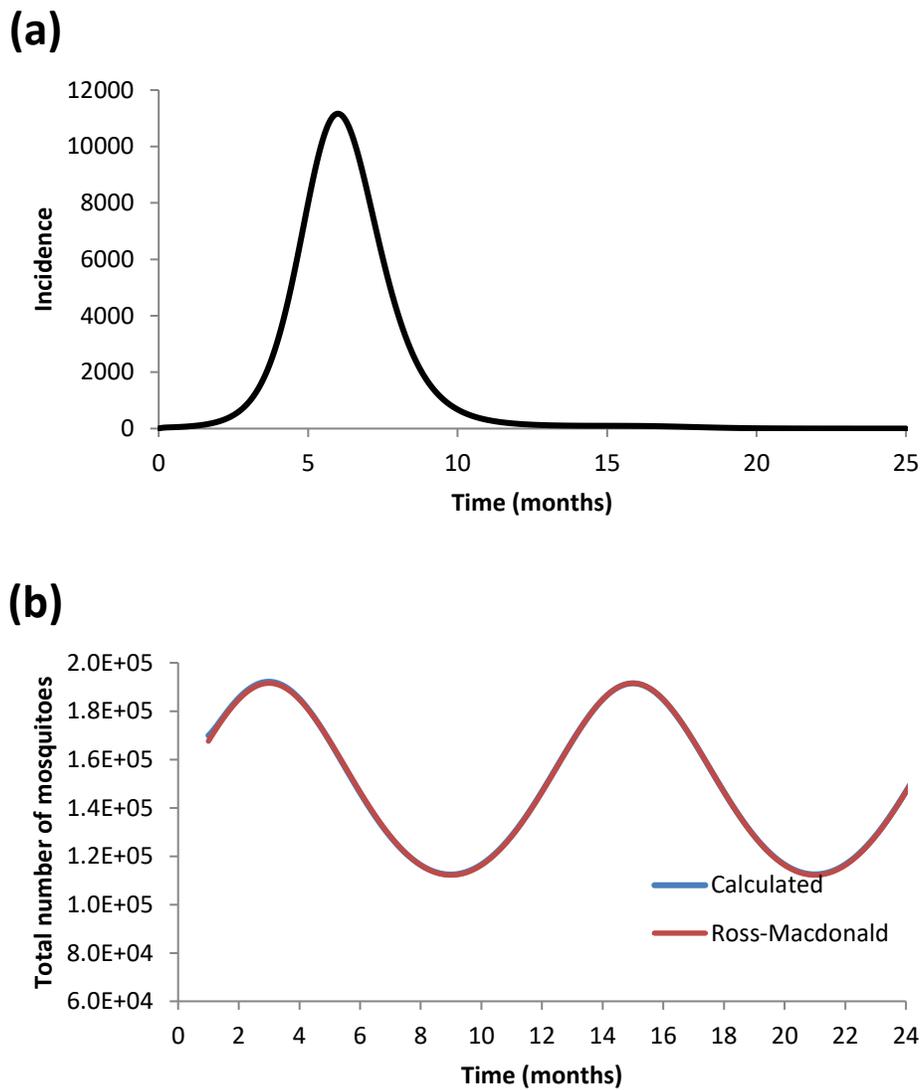

Fig. 6. (a) Dengue incidence outbreak constructed with a seasonal mosquito population. (b) Calculated number of mosquito populations (blue line) compared with that generated by a Ross-Macdonald model assuming a seasonalmosquito population (red line).

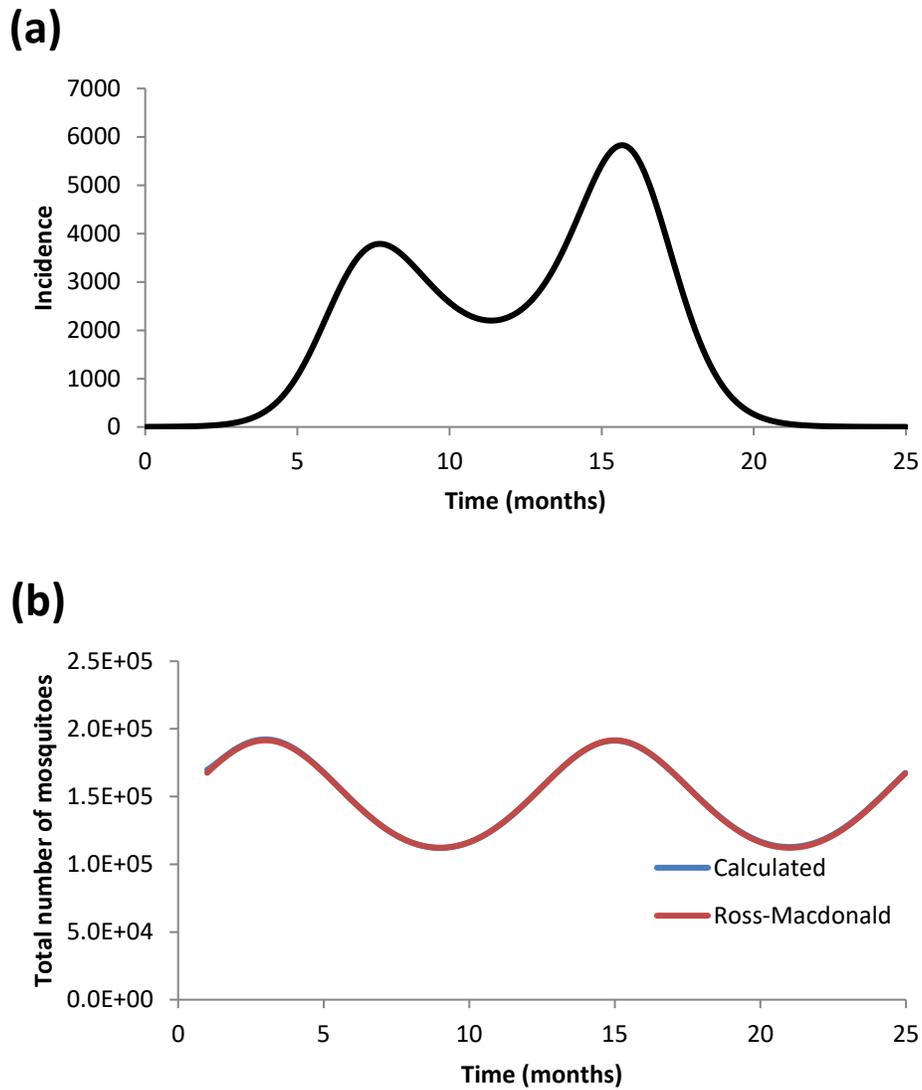

**Fig. 7.** (a) Dengue incidence outbreak constructed with the same seasonal mosquito population as in Fig. 6 but with a different initial condition for the infected humans. (b) Calculated number of mosquito populations (blue line) compared with that generated by a Ross-Macdonald model assuming a seasonalmosquito population (red line).

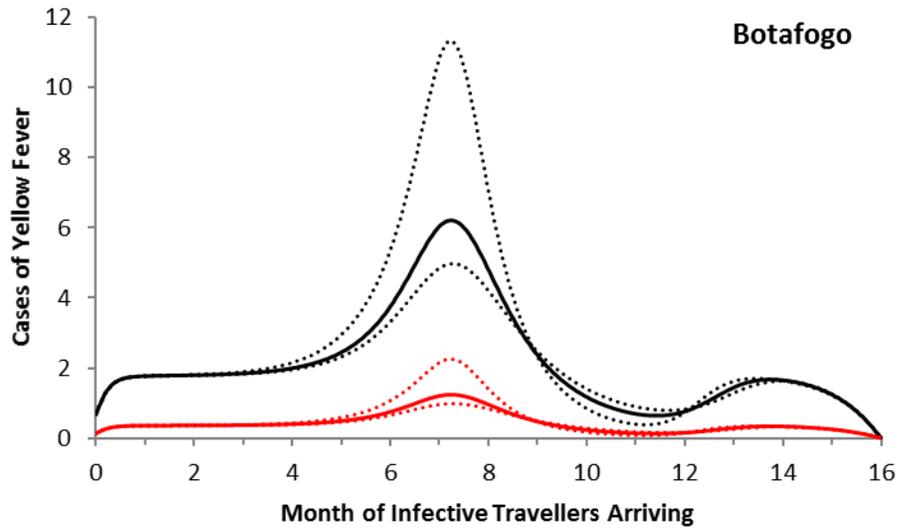

**Fig.8. Total cases of yellow fever in the neighborhood of Botafogo (black line) and symptomatic cases (red line). Dotted lines represent the 95% confidence interval.**

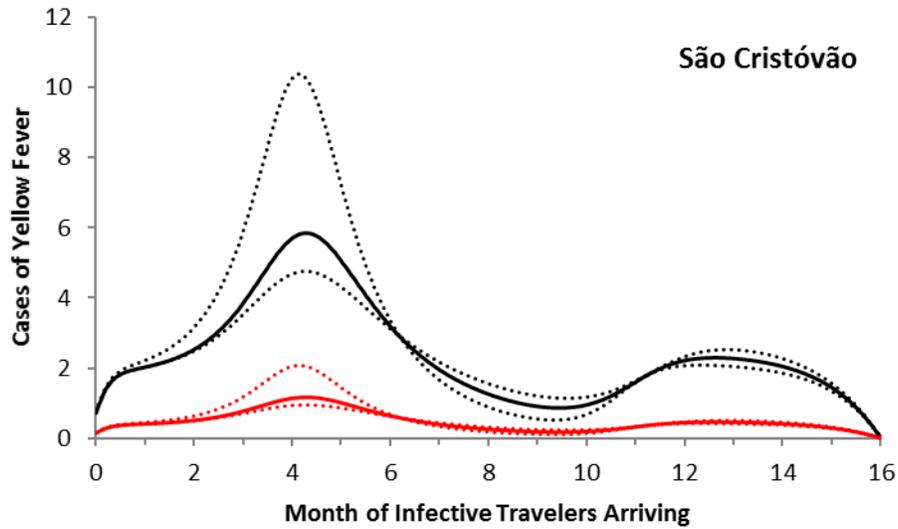

**Fig.9. Total cases of yellow fever in the neighborhood of São Cristóvão (black line) and symptomatic cases (red line). Dotted lines represent the 95% confidence interval.**

**Table 1.**

Model variables and their biological meanings.

| Variable | Biological Meaning |
|---|---|
| $S_H$ | Density of susceptible humans |
| $I_H$ | Density of infected humans |
| $R_H$ | Density of recovered humans |
| $S_M$ | Density of uninfected mosquitoes |
| $L_M$ | Density of latent mosquitoes |
| $I_M$ | Density of infected mosquitoes |

**Table 2.**

Model parameters, their biological meanings and values used.

| Parameter | Meaning | Value | |
|---|---|---|---|
| | | **Dengue** | **Yellow Fever** |
| $a$ | Average daily rate of biting | 10 month$^{-1}$ | 10 month$^{-1}$ |
| $b/b_{yf}$ | Fraction of bites actually infective | 0.6 | variable |
| $\mu_H$ | Human natural mortality rate | 1.19x10$^{-3}$ month$^{-1}$ | 1.19x10$^{-3}$ month$^{-1}$ |
| $\gamma_H/\gamma_{Hyf}$ | Human recovery rate | 4.0 month$^{-1}$ | 6.0 month$^{-1}$ |
| $\gamma_M/\gamma_{Myf}$ | Latency rate in mosquitoes | 5.6 month$^{-1}$ | 4.0 month$^{-1}$ |
| $\mu_M$ | Natural mortality rate of mosquitoes | 5.6 month$^{-1}$ | 5.6 month$^{-1}$ |
| $c/c_{yf}$ | Dengue susceptibility of A. aegypti | 0.6 | variable |
| $\alpha_H/\alpha_{Hyf}$ | Disease-induced mortality rate | 0 | 0.0244 month$^{-1}$ |